**Title:** Characterization of high frequency oscillations and EEG frequency spectra using the damped-oscillator oscillator detector (DOOD)

**Authors:** David Hsu[1]; Murielle Hsu[1]; Heidi L. Grabenstatter[2]; Gregory A. Worrell[3]; Thomas P. Sutula[1]

**Institutions:**

1. Department of Neurology, University of Wisconsin, Madison, WI, United States.
2. Department of Pediatrics, University of Colorado, Aurora, CO, United States.
3. Department of Neurology, Mayo Clinic, Rochester, MN, United States.

**Corresponding author:**

David Hsu

Department of Neurology, Rm 7178

1685 Highland Av

Madison WI 53705-2281

P: 608-265-7951

F: 608-263-0412

E: hsu@neurology.wisc.edu

**Funding**: NIH R01-25020 and NIH R01-NS63039-01.

**Highlights:**

- A novel method called the damped-oscillator oscillator detector (DOOD) has been developed that is capable of very high resolution time-frequency analysis.
- DOOD can be adapted for the reliable, automated detection of high frequency oscillations (HFOs) in the electroencephalogram.
- HFOs as detected by DOOD are associated with intricate low frequency structure.

**Abstract**

*Objective:*  The surgical resection of brain areas with high rates of visually identified high frequency oscillations (HFOs) on EEG has been correlated with improved seizure control. However, it can be difficult to distinguish normal from pathological HFOs, and the visual detection of HFOs is very time-intensive.  An automated algorithm for detecting HFOs and for wide-band spectral analysis is desirable.

*Methods:*  The damped-oscillator oscillator detector (DOOD) is adapted for HFO detection, and tested on recordings from one rat and one human. The rat data consist of recordings from the hippocampus just prior to induction of status epilepticus, and again 6 weeks after induction, after the rat is epileptic.  The human data are temporal lobe depth electrode recordings from a patient who underwent pre-surgical evaluation.

*Results:*  Sensitivities and positive predictive values are presented which depend on specifying a threshold value for HFO detection.  Wide-band time-frequency and HFO-associated frequency spectra are also presented. In the rat data, four high frequency bands are identified at 80-250 Hz, 250-500 Hz, 600-900 Hz and 1000-3000 Hz.  The human data was low-passed filtered at 1000 Hz and showed HFO-associated bands at 15 Hz, 85 Hz, 400 Hz and 700 Hz.

*Conclusion:*  The DOOD algorithm is capable of high resolution time-frequency spectra, and it can be adapted to detect HFOs with high positive predictive value.  HFO-associated wide-band data show intricate low-frequency structure.

*Significance:*  DOOD may ease the labor intensity of HFO detection.  DOOD wide-band analysis may in future help distinguish normal from pathological HFOs.

## Introduction

Detecting transient high frequency oscillations (HFOs) in the human electroencephalogram (EEG) in the ripple (80-200 Hz) and fast ripple (250-500 Hz) range (Bragin, Engel et al. 1999) has attracted increasing interest because ripples and fast ripples are more abundant in seizure onset zones (Staba, Wilson et al. 2002; Worrell, Parish et al. 2004; Staba, Frighetto et al. 2007; Urrestarazu, Chander et al. 2007; Jacobs, LeVan et al. 2008; Jacobs, Levan et al. 2009; Khosravani, Mehrotra et al. 2009; Zijlmans, Jacobs et al. 2009; Bragin, Engel et al. 2010; Jacobs, Zijlmans et al. 2010), and the resection of brain tissue with high rates of ripple and fast ripple activity is correlated with better seizure outcomes (Jacobs, Zijlmans et al. 2010; Wu, Sankar et al. 2010). However, HFOs are also found in normal brain (Buzsaki, Horvath et al. 1992; Ylinen, Bragin et al. 1995; Chrobak and Buzsaki 1996; Draguhn, Traub et al. 2000; Ponomarenko, Korotkova et al. 2003; Axmacher, Elger et al. 2008), and can be difficult to distinguish from normal fast brain oscillations (Engel, Bragin et al. 2009; Nagasawa, Juhasz et al. 2011).

Visual inspection remains the gold standard for the detection of HFOs. The detection of transient oscillatory activity typically involves either band pass or high pass filtering to remove lower frequency activity. For instance, one may employ an 80 Hz high pass filter to detect ripples and a 250 Hz high pass filter to detect fast ripples, with events identified as one or the other if there are at least 4 consecutive oscillations and if the event is clearly above baseline (Zijlmans, Jacobs et al. 2010). Such visual approaches are very time-intensive (Urrestarazu, Chander et al. 2007). A certain amount of subjectivity is also unavoidable.

One early automated approach employs first a band pass filter that passes frequencies between 100 and 500 Hz, then calculating root mean square (RMS) amplitudes of sliding 3 ms time windows of EEG data (Staba, Wilson et al. 2002). Events with peak RMS amplitude greater than 6 standard deviations above the mean and lasting at least 6 ms in duration are identified as candidate HFOs.  A sensitivity of greater than 84% was reported for this method, but the false positive rate was not reported.  The false positive rate in another automated method, the line-length measure, was reported at about 80% when compared with expert visual review (Worrell, Gardner et al. 2008), but sensitivity was not reported.  An earlier test of the line length measure showed that it is superior to the root mean square amplitude measure when tested on EEG data bandpass filtered between 30 and 85 Hz (Gardner, Worrell et al. 2007).  More recently an unsupervised method for classification of high frequency oscillations has been proposed (Blanco, Stead et al. 2010).

As reflected by the variety of approaches for detecting HFOs, the definition of an HFO is somewhat arbitrary, and the various methods that have been applied yield results with a range of sensitivities and specificities that have implications for distinguishing HFOs in pathological tissue from similar oscillating activity in normal tissue.  An automated computer algorithm that detects pathological HFOs quantitatively and reliably would be useful for grading long-term EEG from patients undergoing pre-surgical localization.

We have recently presented a novel pseudo-wavelet approach for high resolution time-frequency analysis called the damped-oscillator oscillator detector (DOOD) (Hsu, Hsu et al. 2010).  This method uses a set of mathematical oscillators, i.e., damped harmonic oscillators, to detect oscillations in EEG. Initial testing suggests that DOOD is capable of superior time-frequency resolution compared with time-windowed Fast Fourier Transform (FFT) approaches and with other wavelet and pseudo-wavelet approaches. Here we show how the DOOD algorithm may be used to detect transient oscillations such as HFOs. After a transient oscillation is detected, one can also investigate the wide-band time-frequency structure surrounding that oscillation. We present illustrative examples from human temporal lobe depth recordings(Worrell, Gardner et al. 2008), and from *in vivo* hippocampal rat recordings.

**Methods**

*Rat data:* Adult Spraque-Dawley rats were placed in a stereotaxic apparatus and anesthetized with 2% isoflurane after pretreatment with atropine (2 mg/kg). The area of incision was injected with 0.5 ml of 0.5% bupivacaine for prolonged local anesthesia/analgesia and postoperative analgesia was administered once daily (buprenorphrine 0.01 mg/kg, subcutaneous). Burr holes for the recording probe and ground screw were placed by conventional surgical techniques. The depth-recording electrode, a 16-channel silicon probe (NeuroNexus, Inc., Ann Arbor, MI) implanted in the right hippocampus (3.0 mm posterior, 2.6 mm lateral, 3.1 mm ventral) spanned CA1, dorsal dentate gyrus, and the hilus at 0.1 mm intervals. Each electrode contact has an area of $7.03 \times 10^{-3}$ mm$^2$. A skull screw was placed in the left frontal bone to serve as ground. Dental acrylic was used to secure the electrodes according to standard chronic methods. At two weeks following surgery for electrode implantation, kainic acid (Tocris Bioscience, Ellisville, MO) was administered in repeated, low doses (5 mg/kg, intraperitoneal) hourly until each rat experiences convulsive status epilepticus for >3 h (Hellier, Patrylo et al. 1998).

Simultaneous *in vivo* field potential recordings from 16-channel silicon probes were obtained in freely-moving rats on a weekly basis for six weeks. One to two minutes of artifact free EEG was saved from each recording session. Field potentials were recorded at a sampling frequency of 12,207.03 Hz and filtered on-line between 2 Hz and 6000 Hz using a Tucker Davis Technologies recording system (Alachua, FL). Off-line conversion of the electrographic data to f32 binary files was accomplished using custom-written software developed in MATLAB® (The MathWorks, Inc., Natick, MA).

The EEG data used for the current study consists of the first 100 seconds from the first rat at baseline (just prior to kainate-induced status epilepticus), and the last 100 seconds from the same rat 46 days later.

After completion of long-term recordings epileptic rats were perfused with 0.01M phosphate buffered saline followed by an aqueous solution of aldehyde fixatives (4% paraformaldehyde-0.5% glutaraldehyde). The brains were removed, post-fixed overnight in same fixative solution, and sectioned on a Pelco Vibratome 1500 sectioning system into 100-μm coronal sections. The sections were wet-mounted in 0.9% saline and imaged with a digital camera (Spot II; Diagnostic Instruments, Sterling Heights, MI) on a Nikon E600 Eclipse epifluorescent microscope x1-4 planapochromatic objectives in order to locate and measure the recording tracts. Brightfield images were acquired at an initial 36-bit tone scale and saved as 16-bit files. The final images were prepared in Adobe Photoshop 7.0. Adjustments in tone scale, gamma, contrast, and hue and subsequent sharpening with the unsharp mask algorithm were applied to the entire image.

The 16 recording sites along the probe length were reconstructed and associated with relevant hippocampal structures using histological measurements and the known inter-site intervals. Electrodes 2-3 were in stratum oriens, with electrode 1 probably in alveus. Electrodes 4-5 were in the CA1 pyramidal cell layer. Electrodes 6-7 were in stratum radiatum. Electrodes 8-9 were in the lacunosum moleculare layer. Electrodes 10-11 were in the molecular layer of the dentate gyrus. Electrodes 12-13 were in the dentate granule cell layer. Electrodes 14-16 were in the hilus of the dentate gyrus.

Institutional regulations for the care of all animals were strictly adhered to, and approval of protocols was obtained from the animal care committee. Animals that are persistently ill after induction of seizures or that suffer frequent, prolonged seizures are euthanized.

*Human depth recordings with microwires.* After Institutional Review Board approval and with informed patient consent, EEG data were collected from a patient with refractory temporal lobe epilepsy who underwent pre-surgical temporal lobe depth electrode evaluation (Worrell, Gardner et al. 2008). Thirty minutes of artifact free stages 3 and 4 slow wave sleep were acquired using custom hybrid depth electrodes which combine both clinical macroelectrodes and special microwires (Adtech Inc.). Each microwire has a contact area of about $1 \times 10^{-3}$ mm$^2$. EEG data were sampled at 5 KHz and low pass filtered at 1000 Hz (Neuralynx, Inc.). The first 120 seconds from the first 8 microwire channels of this data were subsequently used for analysis.

*Visual identification of HFOs.* EEG data were high pass filtered at 80 Hz and visually inspected second-by-second at consecutive one-second intervals for oscillatory activity. An oscillatory event with at least 3 crests or 3 troughs was marked as a putative

HFO if it clearly stood out from baseline. This definition is less restrictive than that used by most investigators. Our motivation is that we wish to include as many potential oscillations as possible, and then to test the ability of DOOD to detect both the ambiguous oscillations and the unequivocal ones depending on a threshold value intrinsic to the DOOD algorithm. In practice, we found the EEG data from the normal rat to contain the most challenging visual samples of putative oscillations, as there were many oscillations that almost blended into the background, when compared both to the sample from the human subject with epilepsy and to the sample from the same rat 6 weeks after kainate-induced status epilepticus.

*Basic DOOD algorithm.* There are several useful variations of the DOOD algorithm (Hsu, Hsu et al. 2010). Let $x_{data}(k,t)$ denote the voltage measured by EEG channel $k$ at time $t$, with $k = 1$ to $N_C$. In the coordinate or *X-DOOD* variation, the data driving force from channel $k$ at time $t$ is taken to be $h(k,t) = x_{data}(k,t)$. In the velocity or *V-DOOD* variation, the data driving force is taken to be the first time derivative of the EEG output, $h(k,t) = \dot{x}_{data}(k,t)$, where the time derivative is performed by finite difference, $\dot{x}_{data}(k,t) = [x_{data}(k,t+\delta t) - x_{data}(k,t)]/\delta t$. Here $\delta t$ is the sampling time related to the sampling rate by $f_S = 1/\delta t$. The $n^{th}$ mathematical oscillator associated with the $k^{th}$ EEG channel at time $t$ is defined by its coordinate $x(k,n,t)$, velocity $\dot{x}(k,n,t)$, frequency $f(n)$, and frictional damping constant $g(n)$. The total number of mathematical oscillators for each EEG channel is denoted $N_f$. Its pseudo-wavelet wavefunction $\psi(k,n,t)$ is given by:

$$\psi(k,n,t) = \int_0^t dt' \, h(k,t') \exp\{-2\pi[g(n) - if(n)](t - t')\}. \tag{1}$$

Taking $\psi_R(k,n,t)$ and $\psi_I(k,n,t)$ to be the real and imaginary parts of $\psi(k,n,t)$, we can write the coordinate and velocity of the associated mathematical oscillator as:

$$x(k,n;t) = \frac{1}{2\pi f(n)} \psi_I(k,n,t) \tag{2}$$

$$\dot{x}(k,n,t) = \psi_R(k,n,t) - \frac{g(n)}{f(n)} \psi_I(k,n,t) \tag{3}$$

For incrementally small time steps $\delta t$, $\psi(k,n,t+\delta t)$ can be easily calculated from $\psi(k,n;t)$ through the recursion relationship:

$$\psi(k,n,t+\delta t) = h(k,t+\delta t)\delta t + \exp\{-2\pi[g(n) - if(n)]\delta t\} \psi(k,n,t) \tag{4}$$

For both X-DOOD and V-DOOD, the spectral density is defined as that part of the time rate of change of the energy at time $t$ due to the data driving force, which is given by:

$$S(k, f(n), t) = \dot{x}(k,n,t) \, h(k,n,t). \tag{5}$$

In practice, we average $S(k,f,t)$ over consecutive time windows of time duration $t_{win}$, in large part to avoid saving enormous files sampled at the EEG sampling time of $\delta t$. When one wishes to detect events of short duration, one should choose $t_{win}$ to be somewhat smaller than the expected event duration. For the detection of HFOs that last tens of milliseconds, we choose $t_{win} = 5\,\text{ms}$.

A third variation of the DOOD algorithm is obtained if one squares the data power before performing the time average:

$$S^2(k, f(n), t) = \left[\dot{x}(k, n; t)\, h(k, n, t)\right]^2. \tag{6}$$

When Eq. 6 is paired with $h(k,t) = \dot{x}_{data}(k,t)$, we refer to this variation as *V-DOOD-SQR*. The X-DOOD and V-DOOD-SQR variations are best for revealing low frequency time-frequency structure, while the V-DOOD variation is most useful for detecting high frequency activity (Hsu, Hsu et al. 2010). For the purpose of detecting HFOs, we will use V-DOOD only. However, we will also show V-DOOD-SQR time-frequency spectral densities to demonstrate associated low-frequency structure.

The friction constant $g(n)$ is equal to the half-width at half-maximum of the spectral density of the $n^{th}$ mathematical oscillator. Therefore, the relative frequency resolution of the $n^{th}$ mathematical oscillator is given by the ratio $g(n)/f(n)$. If all the mathematical oscillators are required to have the same relative frequency resolution, then $g(n)$ must be proportional to $f(n)$, i.e., $g(n) = g_0 f(n)$, where $g_0$ is a dimensionless number which is the same for each mathematical oscillator. In addition, in order to obtain good frequency coverage, the spacing of consecutive oscillators along the frequency axis should be no more than $g(n) = g_0 f(n)$. Therefore, $f(n)$ must have the form of a geometric series:

$$f(n+1) = (1 + \lambda g_0) f(n), \tag{7}$$

where the constant $\lambda$ is a number that is less than or equal to 1. In our experience, choices of $g_0 = 0.02$ to $0.10$ and $\lambda = 0.5$ to 1 work well.

Spanning the frequency range from 1 to 6000 Hz with $g_0 = 0.02$ and $\lambda = 1$ would require $N_f = 440$ mathematical oscillators. A larger value of $g_0$ is permissible (and computationally cheaper) when one is trying to detect oscillatory activity of shorter duration, because the spectral widths of brief oscillations are broader than those of oscillations that persist for many cycles. If we choose $g_0 = 0.10$ and $\lambda = 0.5$, $N_f = 179$ mathematical oscillators would be required. If one attempts to span the same frequency range with linearly spaced oscillators with a frequency resolution of 1 Hz, one would need 6000 oscillators. A geometric frequency series can span the same frequency range as a linear one but much more efficiently, and with no loss of relative frequency resolution.

Note that the most natural way of plotting a geometric frequency series is as the logarithm of the frequency. For convenience, we adopt the base 10 logarithmic scale, denoted as $f_{Log} \equiv Log(f)$ where $f$ is the frequency in Hz. A logarithmic frequency axis allows us to plot wide-band spectra much more conveniently than the more traditional linear frequency axis. It may be of interest that the cochleas of humans, elephants, cats, mice and a number of other animals that are capable of wide-band auditory processing also use a logarithmic frequency scale (Greenwood 1990). In fact, the original motivation for the basic DOOD algorithm came from consideration of a simplified mathematical model of the human cochlea (DH and MH, unpublished).

*HFO detection algorithm.* In using DOOD to detect HFOs, we adopt the following algorithm. First, every EEG channel is individually Z-normalized by subtracting out the mean and dividing by the standard deviation, with the means and standard deviations calculated over the entire data sample. Next, for a given sampling rate $f_S$, we take $f(1) = 1$ Hz, $f_{max} = f(N_f) \approx f_S/2$, $\lambda = 0.5$, and $g_0 = 0.10$. Here $N_f$ is the number of mathematical oscillators for each EEG channel, and the frequency of the $n^{th}$ oscillator is given by Eq (7). The V-DOOD spectral density $S(k,f,t)$ is calculated at each sampling time with sampling time increment $\delta t = 1/f_S$, and then averaged over consecutive time windows of duration $t_{win} = 5$ ms. Let this time-averaged $S(k,f,t)$ be denoted $\langle S(k,f,t) \rangle$. The time resolution of $\langle S(k,f,t) \rangle$ is thus $t_{win} = 5$ ms. The time-averaged spectral density $\langle S(k,f,t) \rangle$ is accumulated over consecutive one-second time intervals. For each one-second time interval and for each individual channel $k$, the means and standard deviations of $\langle S(k,f,t) \rangle$ are calculated with frequencies restricted to a target range of 80 to 1000 Hz. Data points for $\langle S(k,f,t) \rangle$ outside of this frequency range are saved but not used in calculating means and standard deviations. These means and standard deviations are then used to Z-normalize each one-second segment of $\langle S(k,f,t) \rangle$ over the entire frequency range, $f = 1$ Hz to $f_{max}$.

We will perform two types of searches on the function $\langle S(k,f,t)\rangle$. The first is a frequency domain search for a given fixed point in time with the search conducted within the pre-identified frequency range of 80 to 1000 Hz. The purpose of this search is to identify the largest value of $\langle S(k,f,t)\rangle$ at a fixed point in time $t$, denoted $S_{peak}$. The frequency associated with $S_{peak}$ is denoted $f_{peak}$, and a time period of oscillation $t_{peak} = 1/f_{peak}$ is similarly defined. The second type of search is a time domain search, the purpose of which is to identify onset and end times for the candidate HFO. The time step for this time domain search is $t_{win} = 5$ ms. At every time step of the time domain search, $S_{peak}$ and $f_{peak}$ are calculated and updated if the new value of $S_{peak}$ is larger than the prior.

The onset time of a candidate HFO is identified by searching $\langle S(k,f,t)\rangle$ in the time domain for the first occurrence where $S_{peak} > 1$. The end time of the candidate HFO is defined as the first occurrence after onset where $S_{peak} < 1$, *provided a number of other conditions are met*, as follows. First, $S_{peak}$ must be less than 1 for a time period that is at least one full oscillation cycle ($t_{peak}$) past the last time that $S_{peak} > 1$. This precaution is taken because, in the presence of a well-developed EEG oscillation, the DOOD spectral density exhibits a pronounced "beating" effect that depends on the details of the oscillation and whether the oscillation is coupled to other oscillations (Hsu, Hsu et al. 2010). If the flagged end time is not yet one full cycle past the last time that $S_{peak} > 1$, then one keeps stepping through time until one full cycle is reached before accepting the event as having ended.

Once the candidate HFO has definitively ended, then $\langle S(k,f,t)\rangle$ is time-averaged between the onset and end times to produce the doubly time-averaged $\langle\langle S(k,f,t)\rangle\rangle$. A frequency domain search is then performed on $\langle\langle S(k,f,t)\rangle\rangle$ to identify its maximum value, which is denoted $S^*_{peak}$, and is referred to as the *oscillation amplitude index* for that candidate HFO. If $S^*_{peak}$ does not exceed a threshold $S_0$ (the *oscillation amplitude index threshold*), then the entire event is rejected as a candidate HFO. If $S^*_{peak} > S_0$, on the other hand, then we apply one final test. The purpose of the final test is to exclude high amplitude events that are non-oscillatory, or that are over-damped. To perform this test, a frequency domain search is performed on $\langle\langle S(k,f,t)\rangle\rangle$ to identify not only its maximum value, $S^*_{peak}$, but also its full-width at half-maximum, $W^*_{peak}$. We then require that $W^*_{peak} < S^*_{peak}$.

If an event meets all of the criteria above for an HFO, then it is accepted as a candidate HFO, and $S^*_{peak}$ is saved as the oscillation amplitude index of that candidate HFO. The frequency associated with $S^*_{peak}$ is denoted $f^*_{peak}$. The oscillation amplitude index defined in this way has dimensionless units (denoted *Z*), and represents the number of standard deviations that $S^*_{peak}$ is above the mean.

Note that this algorithm is best suited for identifying the largest amplitude HFO within a chosen frequency band at any given moment in time. If there is more than one HFO present at the same time, and if one wishes to identify all HFOs present simultaneously, then a more sophisticated algorithm would be needed. For instance, one may break the target frequency band of interest from 80-1000 Hz into smaller sub-bands, or one may add additional bands at higher or lower frequency. Then one searches through each sub-band individually for HFOs in that sub-band. In practice, HFOs with frequencies greater than 80 Hz are relatively rare and brief events, and it is very infrequent that more than one HFO is present, per individual channel, at the same time.

*Summary of DOOD algorithm.* The key definitions of the DOOD HFO detection algorithm are the V-DOOD spectral density, $S(k,f,t)$, and the oscillation amplitude index $S^*_{peak}$. The V-DOOD spectral density, $S(k,f,t)$, is defined for every EEG channel $k$ at every instant in time $t$. For each candidate HFO, a value for the oscillation amplitude index $S^*_{peak}$ is calculated. $S^*_{peak}$ is in essence a time average of $S(k,f,t)$ over the time when a particular HFO is present. A value of $S^*_{peak}$ is defined for a given EEG channel at a given time only if there is an HFO in that channel at that moment in time. For a candidate HFO to be accepted as an HFO, $S^*_{peak}$ must exceed a threshold $S_0$, i.e., $S^*_{peak} > S_0$. This threshold may be chosen to be smaller if higher sensitivity is desired, and larger if higher specificity is desired.

**Results**

Figure 1 shows a sample of control rat EEG prior to and after bandpass filtering at 80-1000 Hz. There are HFOs apparent after bandpass filtering in electrode channels 2, 14 and 16 (numbered bottom to top) at time $t = 4.78$ sec. On the right are the V-DOOD and V-DOOD-SQR time-frequency color contour plots for channel 16. The V-DOOD plot is band restricted to the frequency range 80-1000 Hz to show the frequency range over which we search for an HFO. The V-DOOD-SQR plot is a wide-band color contour plot which shows the intricate low frequency structure associated with the HFO.

Figure 2 shows another set of HFOs from the control rat EEG which are more prominent than those of Fig. 1. The HFO in channel 15 has an oscillation amplitude index of 5.80 standard deviations above the mean. There is again intricate low frequency structure associated with the HFO. This structure is similar in both channels 1 and 15.

An expanded view of the low frequency structure in channel 15 is shown in Fig. 3a. There is an HFO with $f_{peak}^* = 184$ Hz at time 12.24 sec. In addition, there is 750 Hz activity at 14.75 sec, which does not meet our criteria for an HFO because it has a bandwidth that is slightly wider than its mean frequency. Note that the 184 Hz HFO and the 750 Hz activity appear to interrupt the theta oscillation, which resumes after time $t = 16$ sec. Also note the prominent "beating" effect seen in the theta oscillation. This beating effect is not an artifact; it is characteristic of well-developed oscillatory rhythms in the DOOD time-frequency spectra and its detailed analysis can reveal information on whether the underlying periodicity is sinusoidal or more spike-like, and whether there are other frequencies to which the dominant frequency is coupled (Hsu, Hsu et al. 2010). In the time domain, the 750 Hz high frequency activity has a very fast "buzzing" kind of appearance in all 16 channels. A magnification of the 750 Hz activity, after high pass filtering at 250 Hz, is shown in Fig. 3b.

*Comparison with standard Fast Fourier Transform.* Figure 4 shows time-frequency spectra for the same HFO from channel 15 as in Fig. 2 but now using standard short-time Fast Fourier Transform (ST-FFT) for time-frequency analysis. In this method, each EEG time series is divided into consecutive time windows of duration $t_{win}$. Each EEG time window of duration $t_{win}$ is then zero-padded with a time window also of duration $t_{win}$. FFT is then performed on each zero-padded time window of duration $2t_{win}$. In Fig. 4a, $t_{win} = 0.084$ sec, while in Fig. 4b, $t_{win} = 0.042$ sec. The low frequency structure is much less apparent with ST-FFT as compared to DOOD.

*Sensitivity and positive predictive value.* A total of 1191 HFOs were visually identified in the rat data spread across 16 channels, and 1076 HFOs were visually identified in the human microwire data spread across the 8 channels. In the automated detection of HFOs, an amplitude index threshold $S_0$ (in units of standard deviations above the mean) needs to be specified, above which an event is considered a candidate HFO. If $S_0$ is small, then the HFO detection algorithm will be sensitive but not very specific. If $S_0$ is large, then the algorithm is more specific but less sensitive. The specificity of rare events tends to be close to one, even when the positive predictive value is low. Because HFOs are rare events, we report here the positive predictive value rather than the specificity. The dependence of sensitivity and positive predictive value of the V-DOOD HFO detection algorithm is shown in Fig. 5 for both the rat and human data. The rat data combines 100 seconds of control recordings with 100 seconds of data from the epileptic rat (6 weeks after kainate-induced status epilepticus). The human data consists of 120 seconds of continuous recording.

*HFO event rate.* The control and epileptic rat data are acquired from the same rat. Is the number of HFO events per unit time different before induced status epilepticus and 6 weeks after? In Fig. 6 we show the HFO event rate for the rat at baseline and 6 weeks after kainate-induced status epilepticus, for a choice of threshold $S_0 = 3$. The HFO event rate is much higher for the epileptic rat than for the rat at baseline (2-tailed paired t-test, t = 0.0002). Note that the event rate in the epileptic rat shows a strongly laminar distribution, which is not unexpected given the strongly laminar distribution of cell types in the hippocampus. The EEG channels are spatially separated by 01. mm intervals and closely approximate the layers of the hippocampus (see the *Methods* section describing the rat data). Note that channels 14-16 are all in the hilus of the dentate gyrus and all of these exhibit high HFO event rates in the epileptic rat. The HFO event rate for the human microwires is also shown in Fig. 6. The number of HFO events is higher in channels 2, 4, 5 and 7 than in the other channels. The spatial relationship between these channels are more difficult to determine because the microwires used in these recordings are arranged as a "spray" that extends from a modified clinical macroelectrode (see the *Methods* section on human depth recordings).

*Total vs HFO-associated spectral density.* The spatial distributions of the V-DOOD spectral density for the control and epileptic rat are shown in Fig. 7. These distributions can be calculated in two ways. The total spectral density is calculated by time-averaging $S(k, f, t)$ over all time points in the data file. The HFO-associated spectral density is calculated by accumulating $S(k, f, t)$ only for those times when a candidate HFO is identified in the frequency range 80-1000 Hz, with $S_0 = 3$, and then averaging over those times only. Note again the laminar distribution of the spectral density in all 4 contour plots.

In the total spectral density for the control rat, there are prominent frequency bands at 7 Hz (the theta oscillation), 10-60 Hz (alpha-beta-gamma), 600 Hz and 2000 Hz.

We have discussed the 600 Hz and 2000 Hz bands earlier (Hsu, Hsu et al. 2010). The theta and gamma bands tend to co-localize in channels 2, 4, 6, 12, 14 and 16, while the 600 Hz and 2000 Hz bands tend to co-localize in channels 1, 3, 5, 7, 9, 11, 13, and 15. The HFO-associated spectral density of the control rat shows that the broad 10-60 Hz band splits into separate alpha, beta and gamma bands. It is also apparent that there are oscillators in the 100-200 Hz frequency range although they are not as prominent as the lower frequency oscillators. The 100-200 Hz oscillators appear to co-activate with a continuum of lower frequency activity spanning the theta, alpha, beta and gamma ranges, but they do not co-activate with higher frequency activity in the 600 Hz and 2000 Hz bands.

The total spectral density of the epileptic rat shown demonstrates a marked degradation of the theta oscillation and a relative increase in the gamma band with an apparent invasion into lower frequencies. The 600 Hz band is no longer visible. The HFO-associated spectral density of the epileptic rat shows that there are HFOs of frequency near 200 Hz which co-localize and co-activate with a continuum of lower frequency activity spanning the theta, alpha, beta and gamma ranges, but they do not co-activate with higher frequency activity in the 2000 Hz band. There are no longer clear divisions of lower frequency activity into separate theta, alpha, beta and gamma bands, but rather these are fused together into a broader band. Note that the 100-200 Hz band is invisible in the total spectral density in either the control or epileptic rat. As expected, the HFO-associated spectral density is much more sensitive to brief, transient oscillations than the total spectral density.

Figure 8 shows the total and HFO-associated spectral densities for the human microwire data. The total spectral density exhibits bands at 20 Hz, 60 Hz and 700 Hz. Channel 5 is unique in having a relative more activity at 20 Hz and 60 Hz and less at 700 Hz. The HFO-associated spectral density shows that channel 5 exhibits peak activity at 15 Hz, 85 Hz and 400 Hz, while channels 2, 4 and 7 exhibit peak activity at 700 Hz. This distribution suggests that the 85 Hz and 400 HFOs (ripples and fast ripples, respectively) tend to co-localize spatially and co-activate temporally with lower frequency activity in the alpha-beta range, but not with higher frequency activity in the 700 Hz range. There is no activity visible in the 2000 Hz range, but this data was low-pass filtered at 1000 Hz.

Figures 7 and 8 show that the HFO-associated spectral density provides a valuable adjunct to the frequency analysis of EEG data. It is capable of revealing the frequency structure of transient oscillations that are not apparent in the total spectral density.

*Amplitude and duration of HFOs.* Figure 9a shows the number of HFO events vs HFO amplitude index for the control and epileptic rat. There are a similar number of HFOs of small amplitude index in the control and epileptic rat, but there are more HFOs of higher amplitude index in the epileptic rat. Figure 9b shows the number of HFO events vs HFO duration for a choice of $S_0 = 1$. From this figure, it is seen that the overall distribution of HFO durations is similar in the control and epileptic rat. However, Figs. 9c and 9d show that there are more HFOs of higher amplitude in the epileptic rat, and these have time durations of less than 0.1 sec. Figures 9e and 9f show that the distributions of HFO amplitude indices vs frequency are similar in the control and epileptic rat, but there are more HFOs of higher amplitude index in the epileptic rat and these are mostly in the frequency range of 80-300 Hz (ripple range).

A summary of Fig. 9 is that HFOs in the control and epileptic rat span a similar time duration and a similar frequency range, but there are more HFOs of higher amplitude index in the epileptic rat, and these tend to be of shorter duration (<0.1 sec) and lower frequency (80-250 Hz, ripple range).

For the human temporal lobe microwire data, Fig. 10 shows plots of the number of HFO events vs HFO amplitude index, the distribution of HFO amplitudes vs duration, and the distribution of HFO amplitudes vs frequency. The presence of very high amplitude HFOs (amplitude index > 6) suggests the presence of epileptogenic tissue. However, as compared to the rat hippocampus, the highest HFO amplitudes in this sample occurs in the frequency range 700-800 Hz. These HFOs have a time duration of 30-60 ms.

**Discussion**

The DOOD algorithm is capable of detecting very brief, high frequency EEG oscillations. If high sensitivity is desired, the amplitude index threshold $S_0$ may be set at 1, while if high positive predictive value is desired, it may be set at 3 or higher. We find that the HFO event rate is markedly increased in the epileptic rat compared to a control rat. Similarly, the HFO-associated spectral density is much more sensitive to brief, transient oscillations than the total spectral density.

The DOOD spectral densities shown in Figs. (7) and (8) suggest that high frequency activity may be classified into 4 types: type 1 encompasses the frequency range 80-250 Hz ( $f_{Log}$ =1.9 to 2.4, i.e., ripples); type 2, 250-500 Hz ( $f_{Log}$ = 2.4 to 2.7, fast ripples); type 3, 500-1000 Hz ( $f_{Log}$ = 2.7 to 3.0, the 600 Hz band), and type 4,

1000-5000 Hz ($f_{Log} = 3.0$ to $3.7$, the 2000 Hz band). HFOs in the 600 Hz range and even into the 2000-2600 Hz range have been described in primary somatosensory cortex in response to peripheral somatosensory stimulation (Cracco and Cracco 1976; Curio, Mackert et al. 1994; Gobbele, Buchner et al. 1998; Curio 2000; Sakura, Terada et al. 2009). Based on our own results, we suggest that brain areas other than primary somatosensory cortex may also be capable of producing high frequency activity of types 3 and 4, and that these events can occur spontaneously. A caveat here is that high frequency activity in the 2000 Hz band appears to be rather complex and is not strictly oscillatory (see Fig. 3b).

In agreement with others (Staba, Wilson et al. 2002; Bragin, Engel et al. 2010; Jacobs, Zijlmans et al. 2010; Jacobs, Zijlmans et al. 2010), we find that HFOs especially of type 1 are more abundant in the epileptic rat compared to the control rat, and are also abundant in the temporal lobe of the human subject with temporal lobe epilepsy. These HFOs can be quantified with the HFO event rate and visualized in the frequency domain using the HFO-associated spectral density. The HFO-associated spectral density is able to reveal the frequency structure of associated lower and higher frequency activity as well as of the HFOs themselves in a way that is not possible with the total spectral density. HFOs tend to be invisible in the total spectral density because they are so rare and brief. Both the HFO event rate and the HFO-associated spectral density may be useful clinical markers of epileptogenic tissue.

It is of interest that HFOs of type 1 tend *not* to co-localize with HFOs of types 3 and 4, either in space or in time. It may be of interest in the future to investigate whether types 3 and 4 HFOs predominate in normal brain tissue, while types 1 and 2 predominate in epileptogenic tissue. If true, then epilepsy surgery should aim to resect brain tissue where types 1 and 2 HFOs predominate, but spare tissue where types 3 and 4 predominate.

There are many other ways to detect HFOs automatically and quantitatively, among them the traditional Fast Fourier Transform (FFT), wavelet transforms, other pseudo-wavelet transforms, and supervised (Firpi, Smart et al. 2007) and unsupervised (Blanco, Stead et al. 2010) learning computer algorithms. The DOOD algorithm is a pseudo-wavelet transform that is unique among all wavelet and other pseudo-wavelet transforms in that it has a simple physical analogy in the form of the damped harmonic oscillator. This physical analogy allows the DOOD spectral density to be defined in terms of the time rate of change of the detecting harmonic oscillator that is specifically due to the driving force represented by the EEG data, and not confounded by any other contribution to the total energy. As a result, DOOD appears to give superior time-frequency resolution compared to other wavelet and pseudo-wavelet transforms (Hsu, Hsu et al. 2010).

*How does DOOD compare with FFT in detecting HFOs?* FFT algorithms are unsurpassed when one only wishes to know what frequencies are present in a long stream of EEG data and when one does not need to know when these frequencies appear and when they disappear. It is very easy, however, for very brief and rare events such as HFOs to be lost in an ocean of other EEG activity, if FFT is applied to an entire data file. To detect HFOs using FFT, it is necessary to break the total EEG data file into shorter time segments, and then to apply FFT to each segment to determine the frequency content of each segment (see Fig. 4). However, such short-time FFT (ST-FFT) algorithms are severely limited by the uncertainty principle, which states that improving time resolution necessarily degrades frequency resolution and vice versa. For instance, if the time segments are of duration $t_{win}$, then the frequency resolution becomes $\delta f = 1/t_{win}$. For the rat data, if we take $t_{win} = 0.084$ sec, for example, then $\delta f = 12$ Hz. At this level of resolution, one cannot hope to study delta or theta range activity. If one tries to improve time resolution by taking $t_{win} = 0.042$ sec, then the frequency resolution further degrades to $\delta f = 24$ Hz. The loss of low frequency resolution using ST-FFT was demonstrated in Fig. 4 which is to be compared to the DOOD results of Fig. 2. Note that the HFO at $t = 12.24$ sec appears more smeared out along the frequency axis when one uses the shorter FFT time window $t_{win} = 0.042$ sec. In contrast, the effective time window used for the DOOD calculations in Fig. 2 was much shorter, at $t_{win} = 0.005$ sec, and yet the DOOD frequency resolution is better. While DOOD is also subject to the uncertainty principle (Hsu, Hsu et al. 2010), its manifestation appears to be much less severe in DOOD than in ST-FFT.

Note that in generating the ST-FFT contours Fig. 4, we actually zero-pad the EEG data to remove artifactual end-effects that arise from truncating the EEG time series. One might suppose that such zero-padding improves the frequency resolution of the ST-FFT time-frequency spectrum thus obtained, but it does not, because zero-padding contains no information about longer timescale phenomena

We have also been struck by the intricate low frequency structure that accompanies HFOs, most apparent on V-DOOD-SQR color contour plots. HFOs never occur in isolation and are always accompanied by activity in other frequency bands, especially at lower frequency bands. Others have suggested that this low frequency structure may help distinguish pathological from normal HFOs (Nagasawa, Juhasz et al. 2011).  The remarkable capability of DOOD to time-resolve low frequency structure makes it ideally suited for this purpose.  We conjecture that each HFO may have its own wide-band "signature," which may then facilitate its classification into normal vs pathological.  Such wide-band analysis of HFOs may be useful not only for clinical purposes (in the development of an EEG biomarker of epileptogenic tissue), but also for basic science purposes, in pointing to a need to explain not only the HFOs themselves, but also the accompanying lower frequency structure.

Computer learning algorithms such as neural network approaches (Firpi, Smart et al. 2007) may eventually prove superior to DOOD in detecting HFOs.  Supervised learning algorithms, however, entail a certain degree of subjectivity in determining which EEG waveforms truly are HFOs and which are not.  The DOOD algorithm, furthermore, is a general purpose time-frequency transform.  It can also be used to define instantaneous phases of oscillation and amplitude-amplitude and phase-amplitude correlation functions (Hsu, Hsu et al. 2010).

One limitation of the current study is that visual identification of HFOs was necessary against which the DOOD algorithm can be compared.  There is subjectivity in visual identification that cannot be entirely eliminated.  Some of the visually identified HFOs with smaller DOOD spectral densities may not be deemed true HFOs by other expert investigators.  More stringent standards for HFOs would likely result in the elimination of HFOs with small DOOD amplitude indices, and thus would likely improve the apparent sensitivity of the DOOD HFO detection algorithm.

Further, our experience has been that some high frequency EEG oscillations are not necessarily of high amplitude when visualized in the time domain, even after appropriate bandpass filtering. They are distinguished from nearby segments of EEG only in being more nearly sinusoidal. These oscillations, if brief, can be exceedingly difficult to identify visually. These oscillations may nonetheless show up as prominent peaks in a DOOD spectrogram because they drive the corresponding resonant DOOD oscillators of similar frequency in a coherent way. In this case, one may argue that the DOOD spectral density may be a more objective measure of the presence or absence of EEG oscillations than visual inspection.

The concept of discrete EEG oscillations that appear and disappear, moreover, may not do justice to the complexity of high resolution time domain EEG. Sometimes EEG oscillations have a gradual onset, and sometimes they have a waxing and waning course. In this case, there is a certain unavoidable degree of arbitrariness when one defines onset and end times for these oscillations. It may be that, for the purpose of identifying epileptogenic tissue, these complexities are of minor importance, but it may also be that, for purposes of understanding the details of brain dynamics, a more nuanced description of EEG oscillations is needed. The basic DOOD time-frequency spectra may help to provide such a more nuance description.

Another limitation of our study is that the data sample is quite small, only 120 seconds of human 8-channel microwire data and 200 seconds of 16-channel wide-band rat local field data. Visual identification of HFOs is very time intensive. It has been estimated by others that it takes between 3 minutes to 3 hours to mark one minute of one channel of EEG for the presence of ripples and fast ripples, depending on how active that channel is (Urrestarazu, Chander et al. 2007). Our experience has been similar. In this regard, it would be helpful to have a comparison EEG database, with HFOs identified consensually by multiple experts in the field, on which to test proposed algorithms for detecting HFOs. Our goals here are limited mainly to demonstrating the DOOD algorithm as a candidate algorithm and suggesting what to expect for sensitivities and positive predictive values when using the DOOD algorithm. Future studies will address much longer EEG recordings, and more widely distributed channels, so that recordings near known epileptogenic regions may be compared with that near normal brain.

**Conclusions**

DOOD may be a useful way to quantify HFOs for use in long-term EEG analysis. In addition, DOOD time-frequency analysis reveals an intricate low frequency structure that accompanies HFOs that may be useful for wide-band characterization of HFOs. In future, such wide-band characterization may help differentiate pathological HFOs from normal brain oscillations.

**Acknowledgements:** We are grateful to Sue Osting for help with the rat histology. Funding was provided by NIH R01-25020 and NIH R01-NS63039-01.

**Figure captions**

Figure 1. Sample HFOs in control rat at time $t = 4.78$ sec. (a) 16 channel EEG from control rat with no bandpass vs time. Each EEG channel is Z-normalized to itself by subtracting out the mean and dividing by the standard deviation. The hash marks between consecutive channels represents 10 standard deviations. (b) Same as (a) but after bandpass filtering at 80-1000 Hz. (c) V-DOOD time-frequency color contour plot for channel 16. The frequency axis (vertical scale) is shown in terms of $f_{Log} = \text{Log}(f[\text{Hz}])$ and it spans the frequency range from 50-1000 Hz ($f_{Log} = 1.7$ to 3). (d) V-DOOD-SQR time-frequency color contour plot for channel 16. The frequency axis (vertical scale) is shown in terms of $f_{Log} = \text{Log}(f[\text{Hz}])$ and it spans the frequency range from 2-1000 Hz ($f_{Log} = 0.25$ to 3).

Figure 2. Sample HFOs in control rat at time $t = 12.25$ sec. (a) 16 channel EEG from control rat with no bandpass vs time. Each EEG channel is Z-normalized to itself. The hash marks between consecutive channels represents 10 standard deviations. (b) Same as (a) but after bandpass filtering at 80-1000 Hz. (c) V-DOOD-SQR time-frequency color contour plot for channel 15. The frequency axis (vertical scale) is shown in terms of $f_{Log} = \text{Log}(f[\text{Hz}])$ and it spans the frequency range from 2-1000 Hz ($f_{Log} = 0.25$ to 3). (d) Same as (c) but for channel 1. Note that the V-DOOD-SQR plots of (c) and (d) are not band-passed. The table on the right lists the corresponding oscillation amplitude indices $S^*_{peak}$ for these HFOs in units of standard deviations above

the mean, along with their associated frequencies $f^*_{\text{peak}}$ in Hz. Blank spaces in the table indicate absence of an HFO in the corresponding EEG channel at that moment in time. An asterisk after a number in the first column indicates that the corresponding time-domain EEG discharge was visually identified as being an HFO.

Figure 3. (a) Expanded view of V-DOOD-SQR time-frequency structure for channel 15 of the control rat. The frequency axis (vertical scale) is shown in terms of $f_{Log} = \text{Log}(f[\text{Hz}])$ and it spans the frequency range from 1 to 3162 Hz ($f_{Log} = 0$ to 3.5). There is an HFO with $f^*_{\text{peak}} = 184$ Hz at time 12.24 sec, diffuse 750 Hz activity at 14.75 sec, as well as a 7 Hz ($f_{Log} = 0.85$) theta oscillation which is interrupted by high frequency activity and resumes after time $t = 16$ secs. The prominent "beating" effect seen in the theta oscillation is not an artifact; it is characteristic of well-developed oscillatory rhythms in the DOOD time-frequency spectra (see text). (b) 16 channel EEG of control rat vs time after bandpass filtering at 80-1000 Hz. In the time domain, the 750 Hz high frequency activity has a very fast "buzzing" kind of appearance in all 16 channels. A magnification of the 750 Hz activity, after high pass filtering at 250 Hz, is shown in Fig. 3b. Each EEG channel is Z-normalized to itself. The hash marks between consecutive channels represents 10 standard deviations.

Figure 4. Time-frequency analysis of an HFO using short-time fast Fourier transform (ST-FFT), frequency vs time. The color intensity represents a Z-score calculated by subtracting out the mean of the FFT spectral density and dividing by the standard deviation. This HFO is the same one shown in Figs. (2) and (3), channel 15. (a) ST-FFT for time windows of duration $t_{win} = 0.084$ sec. Note the HFO at 184 Hz with several satellite peaks visible (the satellite peaks are artifact due to the FFT algorithm). (b) ST-FFT for shorter time windows of duration $t_{win} = 0.042$ sec. Note the blurring of the frequency structure with the shorter time window.

Figure 5. Sensitivities and positive predictive values vs threshold amplitude index, $S_0$. On the left are results from 16 channels of rat intrahippocampal data, which includes 100 seconds of control rat data and 100 seconds from epileptic rat. On the right are results from 8 channels of human microwire data from a depth electrode in the temporal lobe. These results are calculated from 120 seconds of microwire data.

Figure 6. HFO event rates for rat and human data with amplitude index threshold of $S_0 = 3$. The number of HFO events per second is plotted against EEG channel for rat (a) and human microwire (b) data. The rat data comes from a 16-channel shank with electrode contacts that are 0.1 mm apart. Note the increase in HFO event rates after epileptogenesis. Also note the strongly laminar structure.

Figure 7. Total vs HFO-associated spectral density for the control and epileptic rat, EEG channel vs logarithm of the frequency. For the HFO-associated spectral densities, an amplitude index threshold of $S_0 = 3$ is chosen. The total spectral densities are in the top row, with HFO-associated spectral densities in the bottom row. The control rat data are on the left, and the epileptic rat on the right. Spectral density intensities are shown on a color scale representing the Z-score, which is calculated by subtracting out the mean and dividing by the standard deviation.

Figure 8. Total (left) vs HFO-associated (right) spectral densities for the human microwire data, EEG channel vs logarithm of the frequency. For the HFO-associated spectral density, an amplitude index threshold of $S_0 = 3$ is chosen. Spectral density intensities are shown on a color scale representing the Z-score, which is calculated by subtracting out the mean and dividing by the standard deviation.

Figure 9. Amplitude, duration and frequency of HFOs in control and epileptic rat. (a) HFO event rate vs amplitude index (here denoted S), with lowest amplitude index threshold taken to be $S_0 = 1$. (b) HFO event rate vs duration, with amplitude index threshold taken to be $S_0 = 1$. (c) Scatter plot of HFO amplitude index vs HFO duration for the control rat. (d) Scatter plot of HFO amplitude index vs HFO duration for the epileptic rat. (e) Scatter plot of HFO amplitude index vs HFO frequency for the control rat. (f) Scatter plot of HFO amplitude index vs HFO frequency for the epileptic rat.

Figure 10. Amplitude, duration and frequency of HFOs in human microwire data. (a) HFO event rate vs amplitude index (here denoted S), with lowest amplitude index threshold taken to be $S_0 = 1$. (b) Scatter plot of HFO amplitude index vs HFO duration. (c) Scatter plot of HFO amplitude index vs HFO frequency.

Fig 1

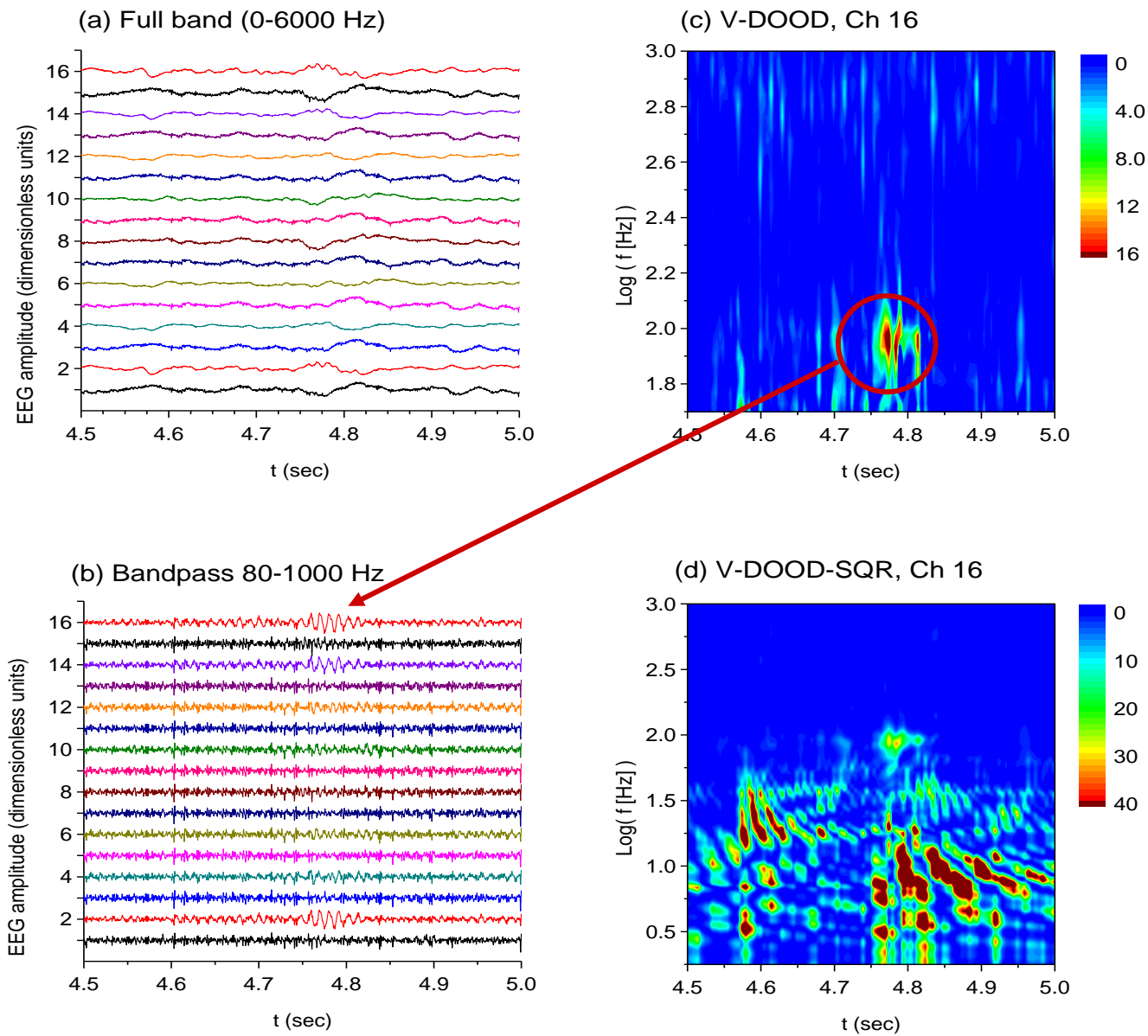

Fig 2

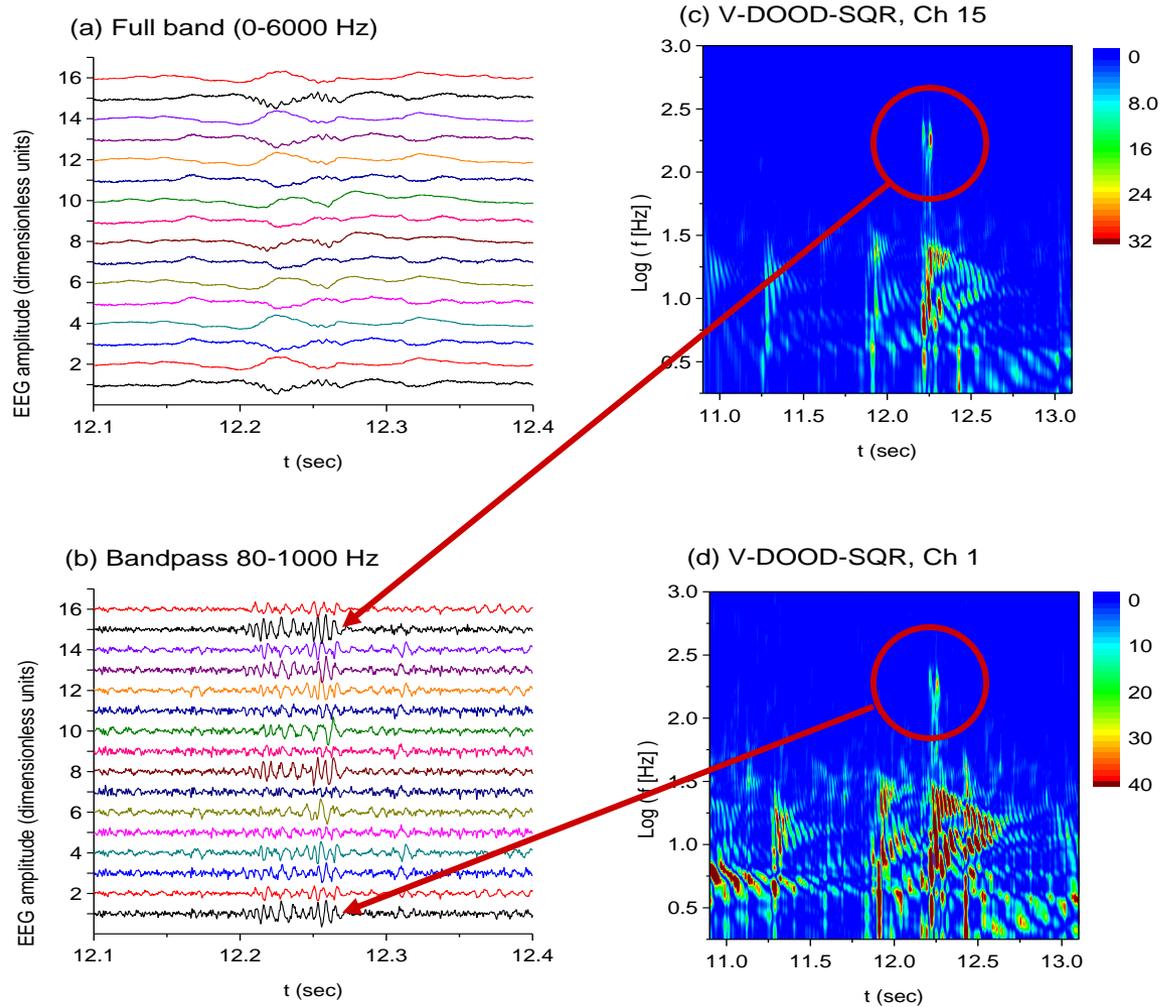

| $S^*_{peak}$ | $f^*_{peak}$ |
|---|---|
| 1.67* | 114 |
| 5.80* | 184 |
| 3.40* | 184 |
| 3.52* | 184 |
| 1.82* | 184 |
| 0.78 | 846 |
| 2.33* | 95 |
| - | - |
| 4.04* | 184 |
| - | - |
| 1.82* | 95 |
| 0.52 | 846 |
| 3.37* | 184 |
| 0.73 | 700 |
| 2.19* | 184 |
| 3.02* | 184 |

Fig 3a

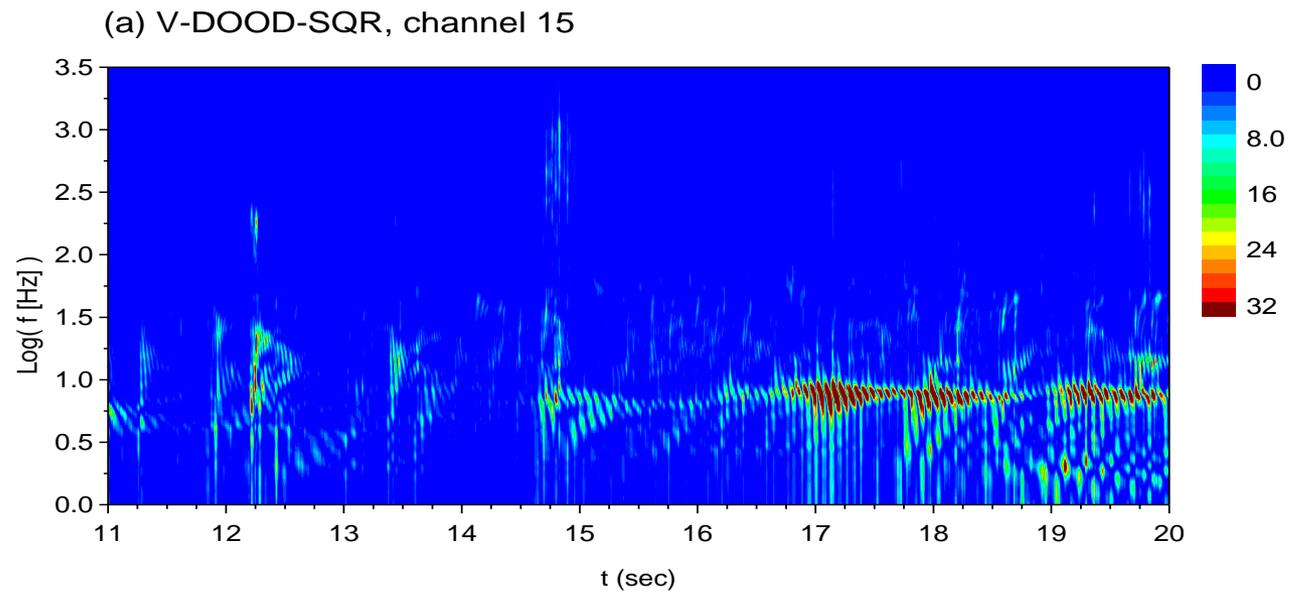

Fig 3b

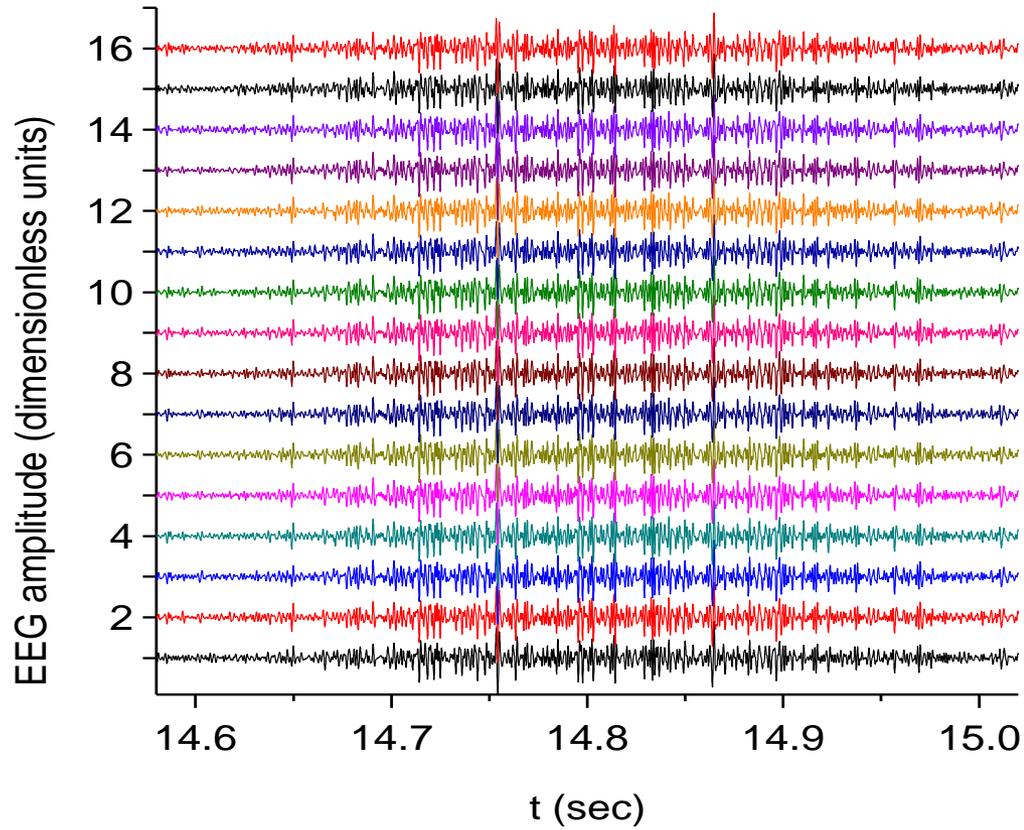

Fig 4

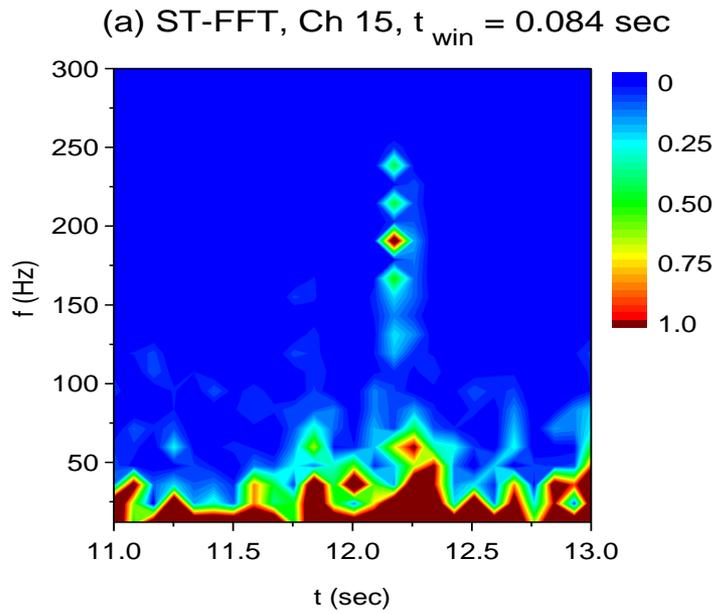 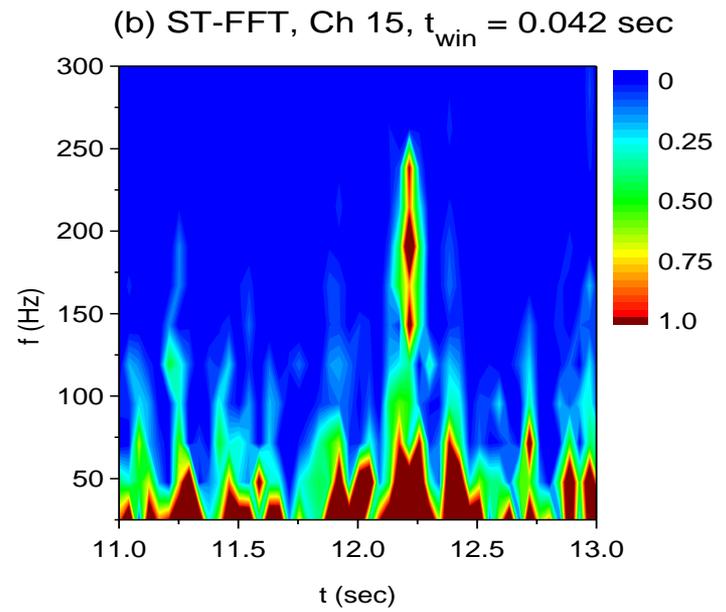

Fig 5

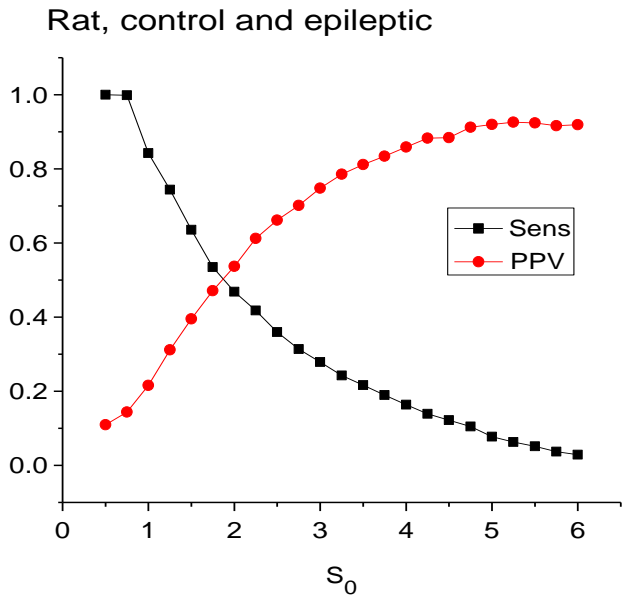
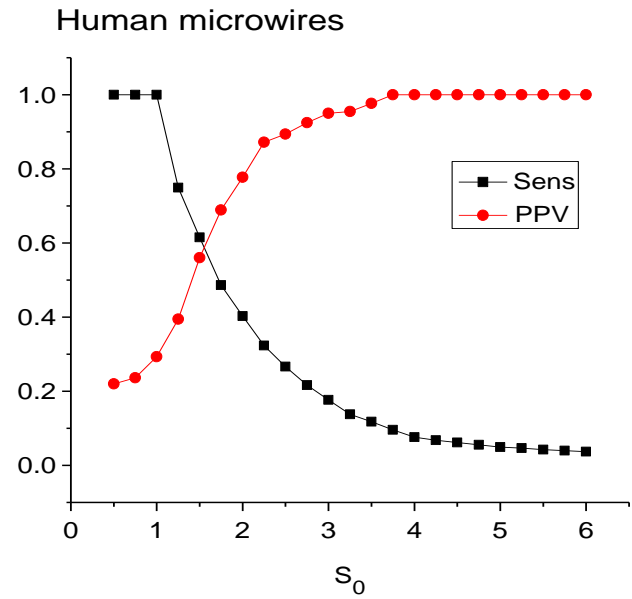

Fig 6

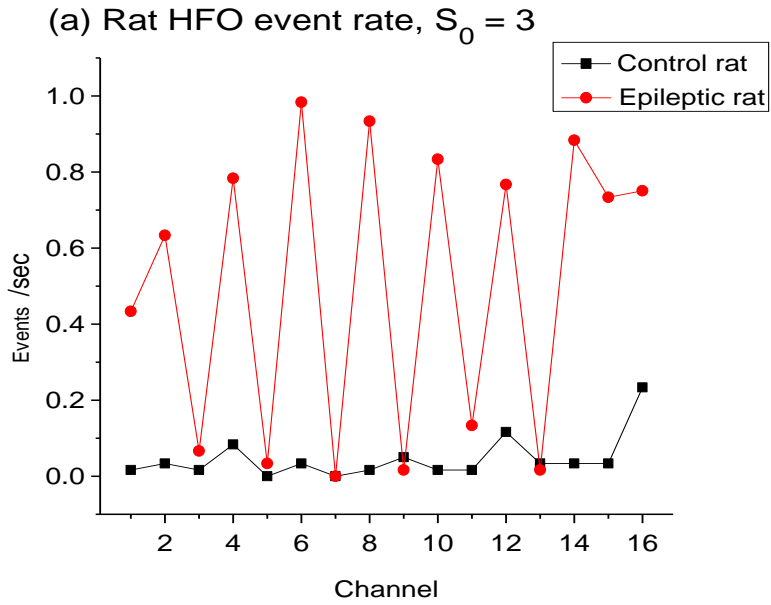
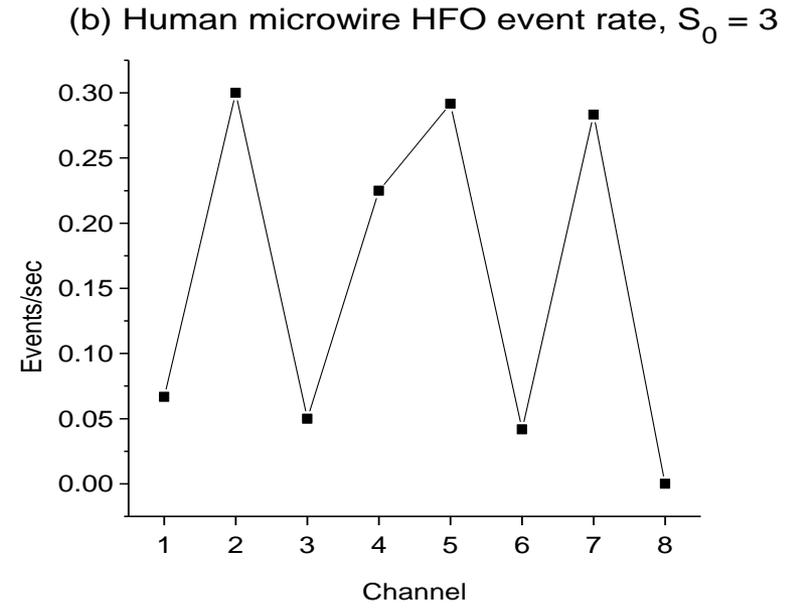

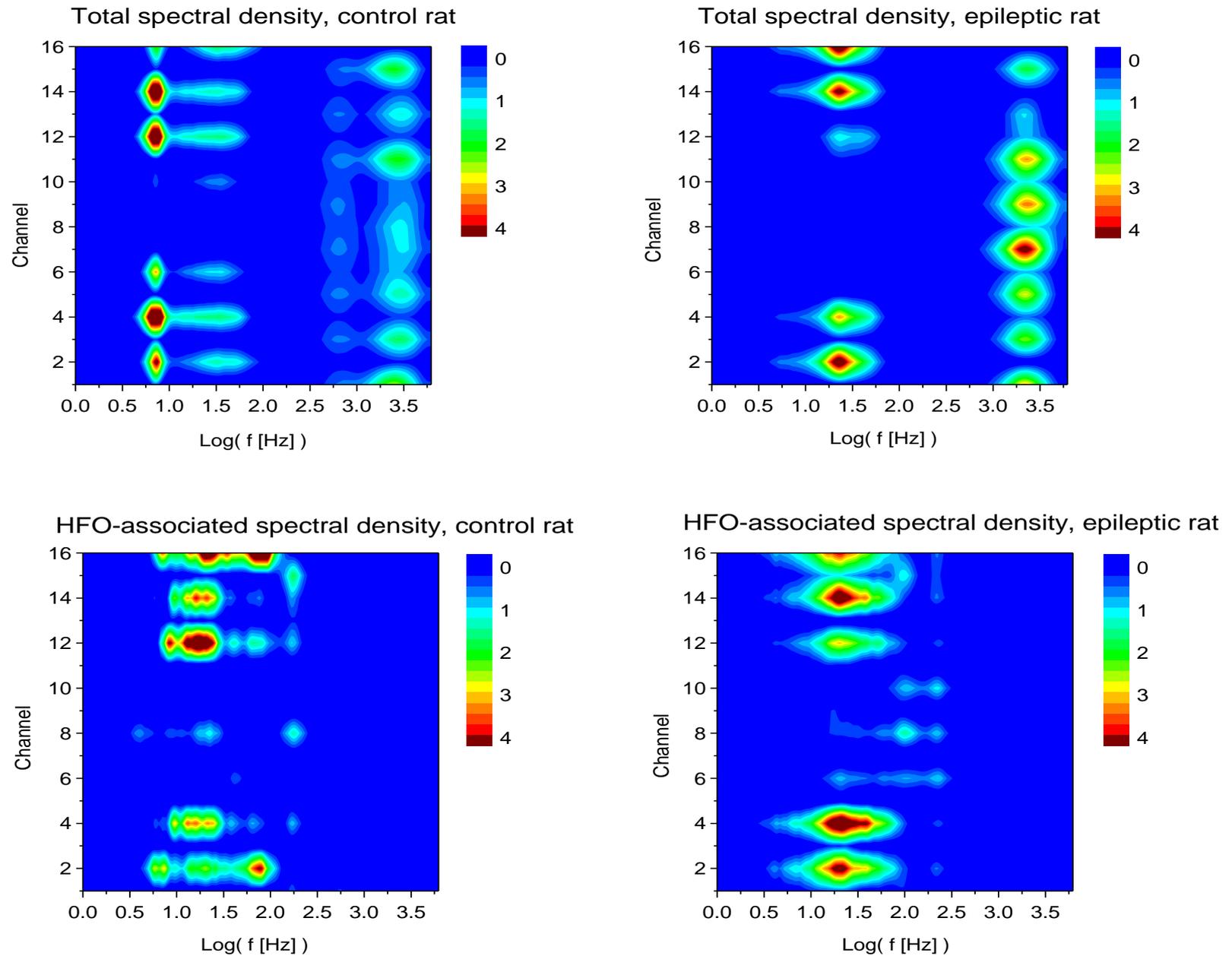

Fig 7

Fig 8

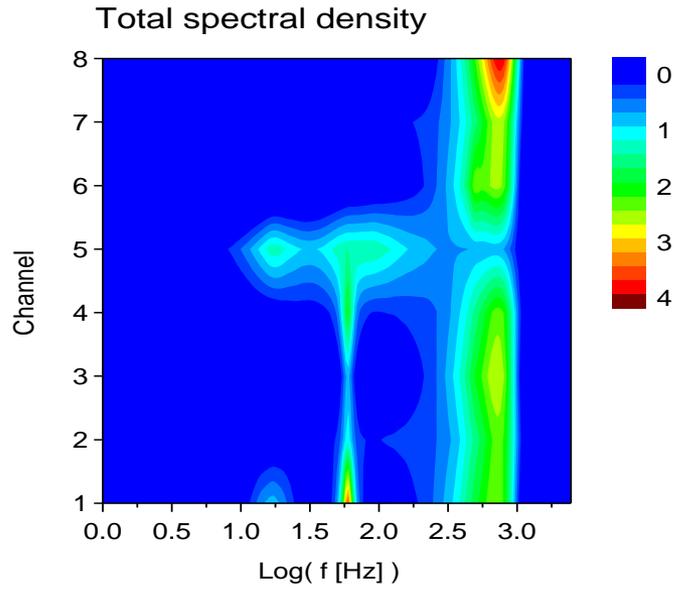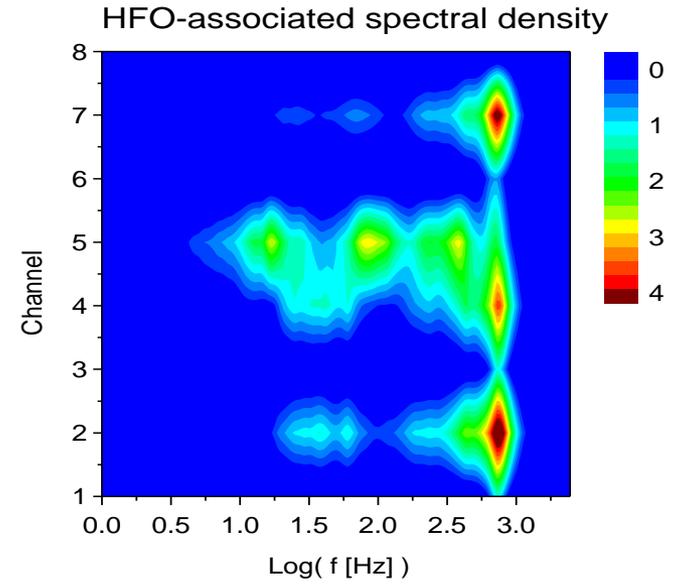

Fig 9

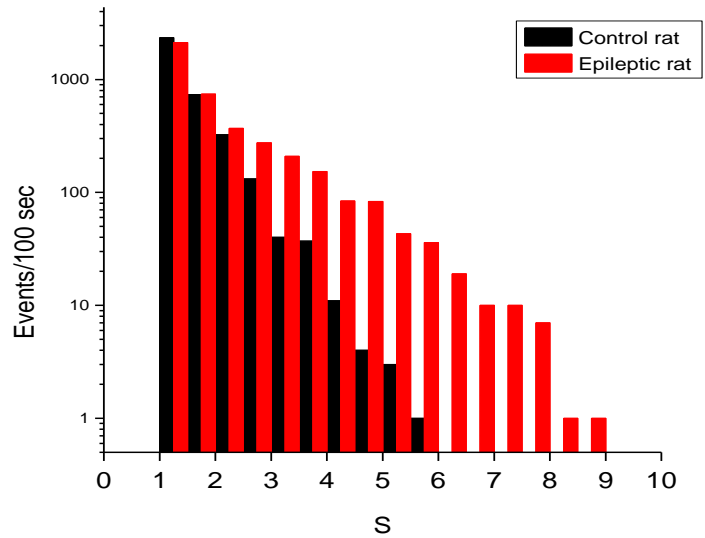
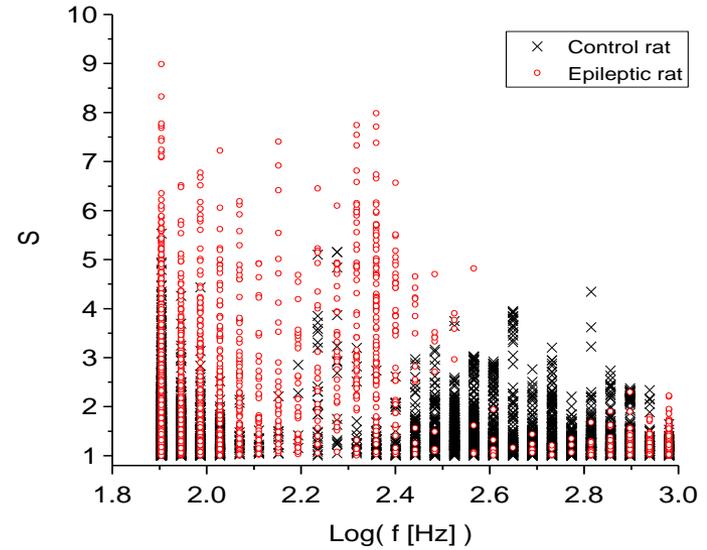
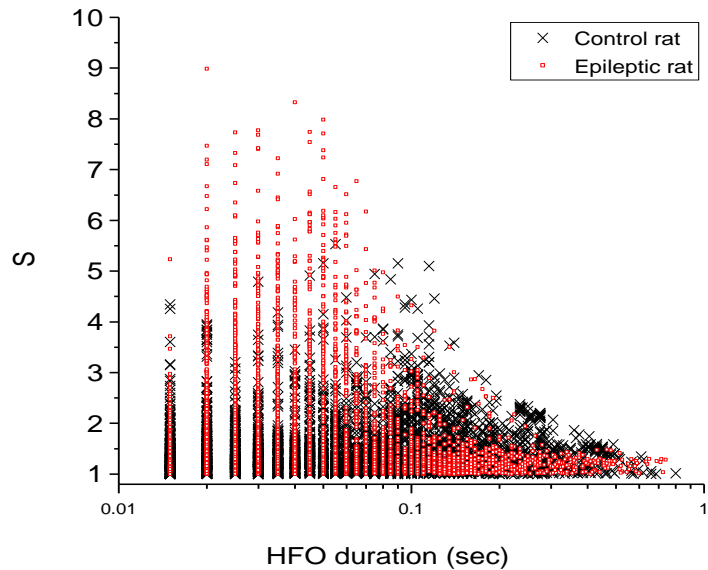
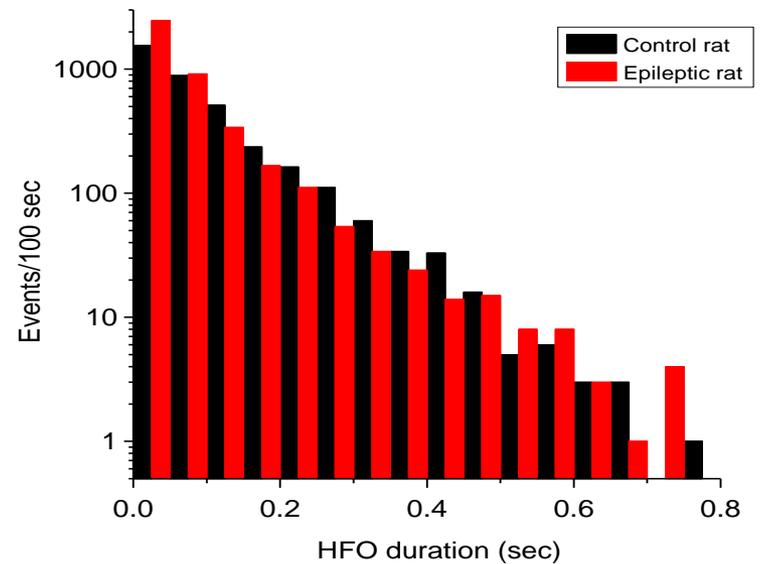

Fig 10a

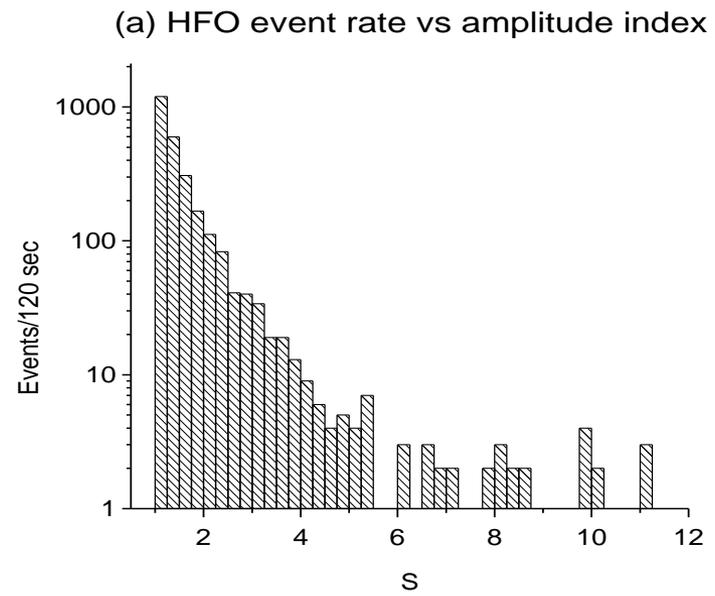

Fig 10b

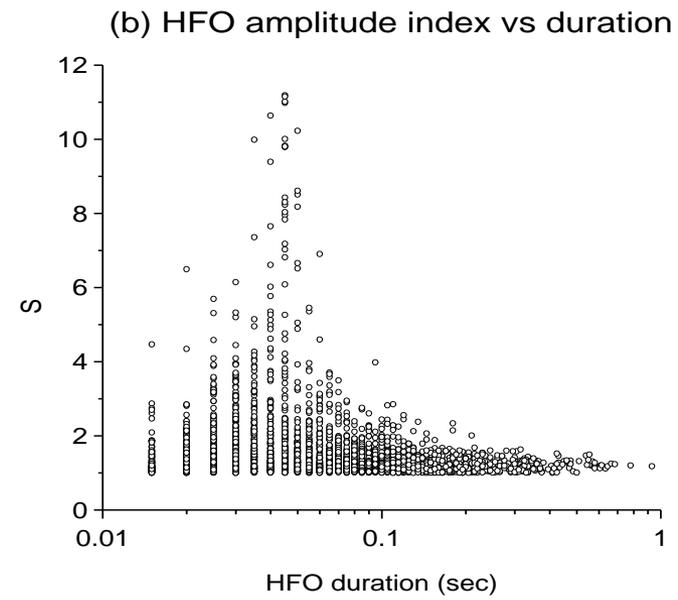

Fig 10c

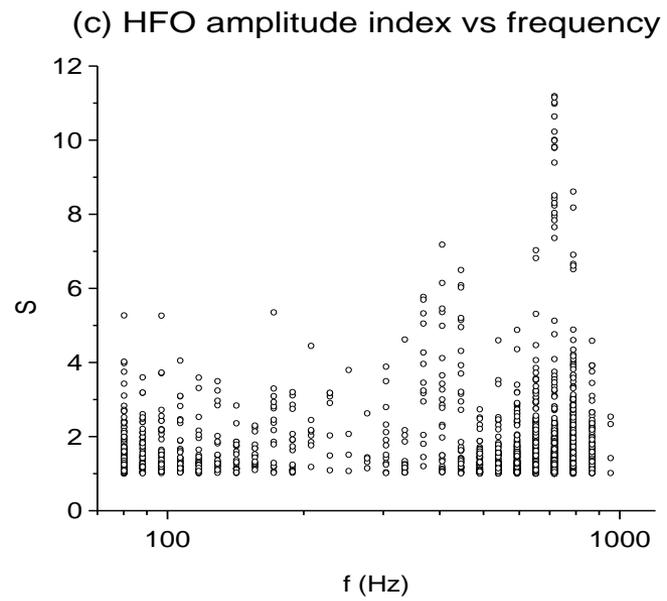